\documentclass{ws-procs9x6}

\newcommand{\req}[1]{(\ref{#1})}

\newcommand{\muF}{\mu_{F}^{2}}
\newcommand{\muO}{\mu_{0}^{2}}
\newcommand{\eps}{\epsilon}
\newcommand{\etaS}{\widetilde{\eta}_1}
\newcommand{\etaO}{\widetilde{\eta}_8}

\def\cbc{{c}\overline{c}}

\begin{document}

\title{Leading-twist two gluon distribution amplitude 
 and  exclusive processes involving $\eta$ and $\eta'$ mesons}

\author{K. PASSEK}

\address{Fachbereich Physik, Universit\"at Wuppertal, 42097
Wuppertal, Germany $^a$\\
E:mail: passek@theorie.physik.uni-wuppertal.de\\[0.1cm]
{\bf (May,  2002)} $^b$
}

\maketitle

\footnotetext[1]{On leave of absence from the 
Rudjer Bo\v{s}kovi\'{c} Institute, Zagreb, Croatia.}
\footnotetext[2]{Talk presented at the workshop on Exclusive Processes at 
High Momentum Transfer, Jefferson Lab, Newport News, Virginia, 15-18 May 2002.} 

\abstracts{
I briefly review the formalism for
treating the leading-twist 
two-gluon states appearing in processes
which involve $\eta$ and $\eta'$ mesons.
The constraints on the size of the lowest order
Gegenbauer coefficients of the two-gluon distribution amplitude
are obtained from the fit to the 
$\eta$ and  $\eta'$ transition form factor data.
The results are applied to
$\chi \rightarrow \eta \eta (\eta' \eta')$ decays
and deeply virtual electroproduction of
$\eta$ and $\eta'$ mesons.
}

\section{Introduction}

The description of the hard exclusive processes 
involving light mesons
is based on the factorization of the short-
and long-distance dynamics and the application
of the perturbative QCD\cite{sHSA}. 
The former is represented by the
process-dependent and perturbatively calculable
elementary hard-scattering amplitude,
in which meson is replaced by its (valent) Fock states,
while the latter is described by 
the process-independent meson distribution
amplitude (DA), 
which encodes the soft physics.
The lowest Fock state of flavour-nonsinglet mesons
consists of quark and antiquark, while
for flavour-singlet meson  the two-gluon state
appears additionally. 
In this paper I give a status report on work of Ref. [2],
and discuss the proper treatment and the importance of 
these two-gluon states.

On the basis of recent results\cite{Leutwyleretc,FeldmannKS98etc}
on $\eta$ and $\eta'$ mixing, we adopt the following
representation of $\eta$ and $\eta'$ in a
octet-singlet basis:
\begin{eqnarray}
|\eta\phantom{'}\rangle &=& \cos{\theta_8}\, |\widetilde{\eta}_8\rangle
                - \sin{\theta_1}\, |\widetilde{\eta}_1\rangle 
  \, ,
       \nonumber \\
|\eta'\rangle &=& \sin{\theta_8}\, |\widetilde{\eta}_8\rangle
                + \cos{\theta_1}\, |\widetilde{\eta}_1\rangle 
  \, ,
\label{eq:OSbasis}
\end{eqnarray}
where the pure octet ($\etaO$) and singlet ($\etaS$)
states are given by
\begin{eqnarray}
\left| \etaS \right> & = &
\frac{f_1}{2 \sqrt{6}} \left[
\phi_1(x) \big| (u \bar{u} + d \bar{d} + s \bar{s})/\sqrt{3} \big>
+ \phi_g(x) \left| gg \right> \right]
\, ,  
\label{eq:FsEta1} \nonumber \\
\left| \etaO \right> &=&
\frac{f_8}{2 \sqrt{6}} \;
\phi_8(x) \big| (u \bar{u} + d \bar{d} - 2  s \bar{s})/\sqrt{6} \big>
\, , 
\label{eq:FsEta8}
\end{eqnarray}
and higher Fock states are neglected.
In this representation, the mixing dependence is solely
embedded in the $\theta_8$ and $\theta_1$ angles, while
in more general approach different distribution amplitudes
$\phi_8^P$ and $\phi_1^P$ could be assumed for $P=\eta,\eta'$.
The numerical values\cite{FeldmannKS98etc}
$f_8 = 1.26 f_\pi$, $f_1 = 1.17 f_\pi$,
$\theta_8 = -21.2^\circ$, and $\theta_1=-9.2^\circ$ are used in this
work.
Alternatively, one could use the recently
suggested quark-flavour basis\cite{FeldmannKS98etc},
but the analysis of DA evolution 
is more straightforward in the above given octet-singlet basis.

\section{Two gluon distribution amplitude and 
the transition form factor for the flavour-singlet meson}

Employing for this section 
more transparent notation $\phi_q \equiv \phi_1$,
the DA evolution equation for $\etaS$ 
takes the matrix form
\begin{equation}
  \muF \frac{\partial}{\partial \muF}
\left(
\begin{array}[c]{c}
\phi_q(x,\muF) \\[0.2cm]
\phi_g(x,\muF) 
\end{array}
\right)=
    V(x,u,\alpha_S(\muF)) \, \otimes \, 
 \left(
\begin{array}[c]{c}
\phi_q(u,\muF) \\[0.2cm]
\phi_g(u,\muF) 
\end{array}
\right)  
\, ,
\label{eq:evDA}
\end{equation}
where $\otimes$ denotes the usual convolution symbol.
The kernel $V$ is 2x2 matrix with a well defined expansion in
$\alpha_S$.
The evolution of the flavour-singlet pseudoscalar meson 
distribution amplitude (DA) 
has been investigated in a number of papers%
\cite{Terentev81,gDAev,Ohrndorf81}.
Most of the results\cite{Terentev81,gDAev} are in agreement
up to differences in conventions. 
On the other hand, the consistent set of
conventions has to be used in calculation of both
the hard-scattering and the distribution amplitude,
and these are not easy to extract from the literature.

Following the recent analysis of the pion
transition form factor\cite{MelicNP01},
we have performed a detailed next-to-leading order (NLO) analysis of
the $\etaS$ transition form factor taking into account
both hard-scattering and perturbatively calculable DA
part, and this enabled us to fix and test the convention
we are using, and to make a connection with other conventions.

The hard-scattering amplitude
one obtains by evaluating 
the $\gamma^* + \gamma \rightarrow q \overline{q}$
and
$\gamma^* + \gamma \rightarrow gg$
amplitudes which we denote by
$T_{q \bar{q}}(u,Q^2)$ and $T_{gg}(u,Q^2)$, respectively.
Owing to the fact that final state quarks and gluons are taken to be
massless and onshell, 
$T_{q \bar{q}}$ and $T_{gg}$ contain collinear singularities,
which have to be factorized out in order to obtain
the finite quantities $T_{H,q \bar{q}}$ and $T_{H,gg}$:
\begin{eqnarray}
T(u,Q^2) &=&
\Big(
T_{H,q \bar{q}}(x,Q^{2},\muF) \quad
T_{H,gg}(x,Q^{2},\muF) \Big) 
\otimes Z^{-1}(x,u,\muF)
\, .
\label{eq:etaT}
\end{eqnarray}
On the other hand, the unrenormalized quark and gluon
distribution amplitudes $\phi_q(u)$ and $\phi_g(u)$
are defined in terms of
$\langle 0 |\bar{\Psi}(-z)  \gamma^+ \gamma_5 \Omega \Psi(z)
| \etaS \rangle$%
and 
$\langle 0 |G^{+ \alpha}(-z) \, \Omega \, 
\tilde{G}^{\: \: \: +}_{\alpha}(z)| \etaS \rangle$, respectively.
The renormalization introduces the mixing of these
quark and gluon composite operators
and
\begin{equation}
\phi(u) =
\left(
\begin{array}[c]{c}
\phi_q(u) \\
\phi_g(u) 
\end{array}
\right) =
Z(u,x,\muF) \otimes
\left(
\begin{array}[c]{c}
\phi_q(x,\muF) \\[0.2cm]
\phi_g(x,\muF) 
\end{array}
\right)
\, .
\label{eq:etaPhiZPhi}
\end{equation}
We note that $Z$ represents a 2x2 matrix.
Perturbation theory cannot be used for a direct evaluation of
$\Phi(u)$, but replacing $|\etaS\rangle$
by $|q \overline{q}\rangle$ or $|gg\rangle$ 
enables us to obtain the perturbatively calculable DA part 
and to determine matrix $Z$.
Finally, the $\etaS$ transition form factor  is
given by 
\begin{equation}
 F_{\gamma^* \gamma \etaS}(Q^{2}) = \frac{f_1}{2 \sqrt{6}} \, 
    T(u,Q^{2})   \, \otimes \,   \phi(u) 
              = \frac{f_1}{2 \sqrt{6}} \,
    T_H(x,Q^{2},\muF)   \, \otimes \,   \phi(u,\muF)
          .
\label{eq:etffCF1}
\end{equation}
Hence, the singularities, which appear in the calculation of 
the hard-scattering  \req{eq:etaT} and  the
DA part \req{eq:etaPhiZPhi}
should cancel, and we have used this requirement to
check the consistency of our calculation. 

By differentiating \req{eq:etaPhiZPhi}
with respect to $\muF$ one obtains
the DA evolution equation \req{eq:evDA} 
with evolution potential $V$ expressed in terms of $Z$
$   V  =  -Z^{-1} \, \otimes \, \left( \muF
       \partial/\partial \muF \,  Z
              \right) $.
The solutions of the leading-order (LO) evolution equation
are given by
\begin{eqnarray}
\phi_q(x,\muF)  &=& 6 x (1-x) 
\left[ 1 
+ \sum_{n=2}^{\infty}{}'
{\  B_n^{q}(\muF)} \:  C_n^{3/2}(2 x -1) \right]
 \nonumber \\[0.1cm] 
\phi_g(x,\muF)  &=&  x^2 (1-x)^2 
 \sum_{n=2}^{\infty}{}'
 { B_n^{g}(\muF)} \:  C_{n-1}^{5/2}(2 x -1)
  \, ,
\label{eq:DAres}
\end{eqnarray}
where
\begin{eqnarray}
{ B_n^q(\muF)} & = &
 B_n^+(\mu_0^2) 
\left( \frac{\alpha_S(\mu_0^2)}{\alpha_S(\muF)} \right)^{%
\frac{ \gamma_{+}^n}{\beta_0}} +
{ \rho_n^-} \,  B_n^-(\mu_0^2) 
\left( \frac{\alpha_S(\mu_0^2)}{\alpha_S(\muF)} \right)^{%
\frac{ \gamma_{-}^n}{\beta_0}} 
 \nonumber \\[0.1cm]
{ B_n^q(\muF)} & = &
 {\rho_n^+} \, B_n^+(\mu_0^2) 
\left( \frac{\alpha_S(\mu_0^2)}{\alpha_S(\muF)} \right)^{%
\frac{\gamma_{+}^n}{\beta_0}} +
 B_n^-(\mu_0^2) 
\left( \frac{\alpha_S(\mu_0^2)}{\alpha_S(\muF)} \right)^{%
\frac{ \gamma_{-}^n}{\beta_0}} 
  \, .
\end{eqnarray}
The coefficients $B_n^{\pm}(\mu_0^2)$, i.e.,
$B_n^{q,(g)}(\mu_0^2)$, represent nonperturbative input at 
scale $\mu_0^2$, while 
$\gamma^{\pm}_n
= 1/2 \left[
\left( \gamma^{qq}_n+ \gamma^{gg}_n \right)
\pm \sqrt{ \left( \gamma^{qq}_n- \gamma^{gg}_n \right)^2
+4 \gamma^{qg}_n \gamma^{gq}_n }
 \right]$
and
\begin{equation}
\rho_{n}^+ = 
    6 \; \frac{\gamma^{gq}_n}{\gamma^+_n - \gamma^{gg}_n}
\, ,
\qquad
\rho_{n}^- = 
    \frac{1}{6}\;  \frac{\gamma^{qg}_n}{\gamma^-_n - \gamma^{qq}_n}
\,  
\label{eq:rho} 
\end{equation}
are defined in terms  of anomalous dimensions
$\gamma^{qq}_n=\gamma_n^{(0)}$\cite{MelicNP01},
$\gamma^{gg}_n$\cite{Terentev81},
\begin{equation}
\gamma^{qg}_n = \sqrt{n_f C_F} \; 
              \frac{n (n+3)}{3 (n+1) (n+2)} 
        \qquad
\gamma^{gq}_n = \sqrt{n_f C_F} \;  
                     \frac{12}{ (n+1) (n+2)} 
     \, . 
\label{eq:andim}
\end{equation}
The DA $\phi_q$ is normalized to 1,
but, since
$ \int_0^1 dx \; \phi_g(x, \muF)=0 $,
there is no such natural way to normalize
$\phi_g$.
It is important to emphasize that any change of the
normalization of the gluon DA
is accompanied by the corresponding change
in the hard-scattering part.
Namely, for $\phi_g \rightarrow \sigma \, \phi_g$,
the projection of $gg$ state on the $\etaS$ state,
which can be derived from the definition of the
gluon distribution amplitude, 
gets modified by factor $1/\sigma$, i.e.,
\begin{equation}
\frac{i}{2}  \; 
\eps^{\mu \nu \alpha \beta} \;
\frac{n_{\alpha} P_{\beta}}{n\cdot P} \;
\frac{1}{x(1-x)}
\rightarrow
\frac{1}{\sigma} \; \frac{i}{2}  \; 
\eps^{\mu \nu \alpha \beta} \;
\frac{n_{\alpha} P_{\beta}}{n\cdot P} \;
\frac{1}{x(1-x)}
\, ,
\label{eq:gg}
\end{equation}
and, thus, the hard-scattering amplitude changes
while the physical quantity ($F_{\gamma* \gamma \etaS}$, in this case)
remains independent of the choice of convention.
By inspecting Eqs. (\ref{eq:DAres}-\ref{eq:rho}),
it is easy to see
that the change of the normalization
of $\phi_g$ can be translated into the change
of the off-diagonal anomalous dimensions 
\begin{equation}
\gamma_n^{qg} \rightarrow \frac{1}{\sigma} \gamma_n^{qg}
\qquad 
\gamma_n^{gq} \rightarrow \sigma  \, \gamma_n^{gq}
\label{eq:anch}
\end{equation}
and
of the coefficient 
$B_n^{-}(\mu_0^2) \rightarrow  \sigma  B_n^{-}(\mu_0^2)$.
The former can be easily understood in the ``operator''
language, i.e., by considering the impact of the change
of the normalization of the gluon operator
on the anomalous dimensions.
Hence, by employing \req{eq:anch} along with \req{eq:andim}
and \req{eq:gg}, we can consistently change our conventions in order to
compare our results with other calculations from the literature.
For historical reasons (comparison with the forward case),
in what follows $\sigma=\sqrt{n_f/C_F}$ is used.

\section{Applications}

\begin{figure}[th]
\centerline{\epsfxsize=3.in\epsfbox{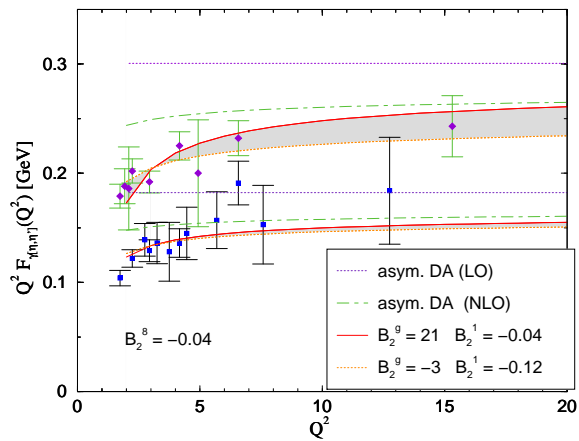}}   
\caption{$\eta$ (below) and $\eta'$ (above) transition form factors. 
The shaded area corresponds to the in Eq. \protect\req{eq:fit81g}
given range for $B_2^1(\muO)$ and $B_2^g(\muO)$.}
\label{f:etff}
\end{figure}

Using the mixing scheme defined in Eq. \req{eq:OSbasis},
we have obtained the NLO leading-twist prediction
for the $\eta$ and $\eta'$ transition form factors.
For the treatment of $\phi_8(x,\muF)$, we use the well-known
LO result for the flavour-nonsinglet meson 
distribution amplitude\cite{sHSA,MelicNP01}.
We truncate the Gegenbauer series at $n=2$, and fit our results
to the experimental data\cite{exptff}. 
The fits are carried through with $\mu_R=Q/\sqrt{2}$ and $\mu_F=Q$,
with $\alpha_S$ evaluated from the two-loop expression with $n_f=4$ and
$\Lambda^{(4)}_{\overline{MS}}=305$ MeV.
For $Q^2\geq 2$ GeV$^2$ and $\mu_0=1$ GeV, the results of the fits read 
\begin{equation}
B_2^8(\muO)  = -0.04 \pm 0.04 \quad 
B_2^1(\muO) = -0.08 \pm 0.04  \quad
B_2^g(\muO)  = 9 \pm 12
\, .
\label{eq:fit81g}
\end{equation}
The existing experimental data and their quality allow us
to obtain not more than a constraint on the value of $B_2^g$.
As expected, we have observed  a strong correlation between $B_2^1$ and $B_2^g$.
The quality of the fit is shown in Fig. \ref{f:etff}.  

As a first application of the above 
given results, we have analyzed the
$\chi_{c0}\to \eta\eta, \, \eta'\eta'$ decays. 
Following previous work\cite{BolzKS96etc},
we take 
$\chi_{c0}$ as a non-relativistic $\cbc$ bound state,
and obtain the decay amplitudes 
for $\cbc \to (q\overline{q})(q\overline{q})$
and $\cbc \to (gg)(gg)$.
The ratio $\Gamma_{\eta' \eta'}/\Gamma_{\eta \eta}$ 
is suitable for investigating the
sensitivity 
to the gluon contributions.
Despite possibly large value of $B_2^g$,
we have observed only modest dependence 
on the variations of $B_1$ and $B_g$ 
in the allowed range \req{eq:fit81g}
(up to 20 \% difference).

Finally, I report on our result for the
two gluon contribution to the deeply virtual
electroproduction of $\eta$, $\eta'$ mesons.
In this case, the subprocess amplitudes
$q + \gamma^*_L \rightarrow (q \overline{q}) + q$
and
$q + \gamma^*_L \rightarrow (gg) + q$
have to be evaluated.
Our result for the former amplitude $H^{P(q)}_{0+,0+}$ is in agreement with
the results from the literature\cite{HuangK00,DVEM}.
In terms of subprocess Mandelstam variables
($\hat{s}$,$\hat{u}$,$\hat{t}=t$), the latter amplitude 
reads
\begin{equation}
H^{P(g)}_{0+,0+}
= \frac{4\pi\alpha_s}{Q} \, \frac{C_F}{N_C}
 \frac{1}{2} \frac{1}{\sqrt{n_f}}
\frac{Q \sqrt{- \hat{u} \hat{s}}}{Q^2+\hat{s}}
\int_0^1 d\tau \frac{\phi_g(\tau)}{\tau (1-\tau)} \; 
           \frac{t}{\hat{u} \hat{s}} \; 
        \left(\frac{1}{\tau}-\frac{1}{1-\tau}\right) 
 .
\label{eq:Hg}
\end{equation}
In deeply virtual electroproduction
of mesons (DVEM), the limit $t \rightarrow 0$ has to
be considered.
For $t=0$ and $\hat{s}+\hat{u}=-Q^2$,
our result \req{eq:Hg} gives $H^{P(g)}_{0+,0+}=  0$.
We conclude that the two-gluon contribution is suppressed.

\section{Conclusions}

We have performed
a detailed analysis of the proper
inclusion of the two gluon states in the 
(octet-singlet scheme based)
description
of the hard exclusive processes involving $\eta$ and $\eta'$
mesons. 
Normalization and conventions have been fixed
and  discrepancies found in the literature have been resolved.
From the fit of the $\eta$, $\eta'$ transition form factors
to experimental data, we have obtained
the range for the 
$B_2^1$, $B_2^g$ and $B_2^8$ coefficients of the
$\phi_1$, $\phi_g$ and $\phi_8$ DAs, respectively. Expected strong
correlation between $B_2^1$ and $B_2^g$ has been observed.
The results have been applied to
$\chi_{c0} \rightarrow \eta \eta (\eta' \eta')$ decay,
where only a moderate dependence on 
$B_2^g$ has been found.
For DVEM, the 
$\gamma^* q \rightarrow (gg) q$ subprocess 
has been found to be suppressed for small
momentum transfer $t$.
We conclude that for considered processes 
the theoretical and experimental
uncertainties do not allow further restriction of the
parameter range since only a modest
dependence on the value of $B_2^g$ has been
observed.
This contrasts the findings 
for $\eta' g^* g^*$ vertex\cite{AliP00}.

\section*{Acknowledgments}
This work is supported by Deutsche Forschungs Gemeinschaft
and partially supported by the Ministry of Science and Tech.,
Croatia (No. 0098002).

\end{document}